\documentclass[aps,prl,twocolumn,nofootinbib,preprintnumbers]{revtex4}

\usepackage{color}

\usepackage[
      colorlinks=true,
      linkcolor=blue,
      urlcolor=blue,
      filecolor=black,
      citecolor=red,
      pdfstartview=FitV,
      pdftitle={},
        pdfauthor={Thomas Faulkner, Hong Liu, Mukund Rangamani},
        pdfsubject={},
        pdfkeywords={},
        pdfpagemode=None,
        bookmarksopen=true
      ]{hyperref}

\usepackage[normalem]{ulem}
\usepackage{amsmath}
\usepackage{enumerate}
\usepackage{amsfonts}
\usepackage{yfonts}

\usepackage{subfigure}
\usepackage{psfrag}

\usepackage{epsfig}
\usepackage[latin1]{inputenc}
\usepackage{float}
\usepackage{graphicx}
\usepackage{cancel}
\usepackage{mathrsfs}
\usepackage{amssymb}
\usepackage{amsfonts}
\usepackage{amsmath}
\usepackage{slashed}
\usepackage{mathtools}

\usepackage{graphicx}
\usepackage{bm}

\def\({\left(}
\def\){\right)}
\def\[{\left[}
\def\]{\right]}
\def\<{\langle}
\def\>{\rangle}

\newcommand\half{{\ensuremath{\frac{1}{2}}}}

\newcommand{\be}{\begin{equation}}
\newcommand{\ee}{\end{equation}}
\newcommand{\bea}{\begin{eqnarray}}
\newcommand{\eea}{\end{eqnarray}}
\newcommand{\bwt}{\begin{widetext}}
\newcommand{\ewt}{\end{widetext}}

\newcommand{\bi}{\begin{itemize}}
\newcommand{\ei}{\end{itemize}}
\newcommand{\ben}{\begin{enumerate}}
\newcommand{\een}{\end{enumerate}}
\newcommand{\bca}{\begin{cases}}
\newcommand{\eca}{\end{cases}}
\newcommand{\bln}{\begin{align}}
\newcommand{\eln}{\end{align}}
\newcommand{\bst}{\begin{split}}
\newcommand{\est}{\end{split}}

\newcommand\Sig{\Sigma}
\newcommand\lam{\lambda}

\newcommand\ga{{\ensuremath{{\gamma}}}}
\newcommand\Ga{{\ensuremath{{\Gamma}}}}
\newcommand\de{{\ensuremath{{\delta}}}}
\newcommand\De{{\ensuremath{{\Delta}}}}

\def\th{{\theta}}

\newcommand\ov{\over}
\newcommand\ha{{\half}}

\def\le{\left}
\def\ri{\right}

\newcommand\sA{{\ensuremath{{\mathcal A}}}}

\newcommand\sE{{\ensuremath{{\mathcal E}}}}

\newcommand\sN{{\ensuremath{{\mathcal N}}}}

\newcommand{\ka}{{\kappa}}

\newcommand{\ft}{\mathfrak{t}}

\newcommand{\fR}{\mathfrak{R}}

\newcommand{\eql}{\ell_{\rm eq}}
\newcommand{\Vee}{v_{E}}

\begin{document}


\title {Entanglement Tsunami: \\
Universal Scaling in Holographic 
Thermalization
}

\preprint{MIT-CTP 4475}

\author{ Hong Liu}
\affiliation{Center for Theoretical Physics,
Massachusetts
Institute of Technology,
Cambridge, MA 02139 }
\author{S.~Josephine Suh}
\affiliation{Center for Theoretical Physics, Massachusetts Institute of Technology,
Cambridge, MA 02139 }

\begin{abstract}

We consider the time evolution of entanglement entropy after a global quench in a strongly coupled
holographic system, whose
subsequent equilibration is described in the gravity dual by the gravitational collapse of a thin shell of matter resulting in a black hole. In the limit of large regions of entanglement, the evolution of entanglement entropy is controlled by the geometry around and inside the event horizon of the black hole, resulting in regimes of pre-local- equilibration quadratic growth (in time), post-local-equilibration linear growth, a late-time regime in which the evolution does not carry any memory of the size and shape of the entangled region, and a saturation regime with critical behavior resembling those in continuous phase transitions. Collectively, these regimes suggest a picture of entanglement growth in which an ``entanglement tsunami''  carries entanglement inward from the boundary. We also make a conjecture on the maximal rate of entanglement growth in relativistic systems.

\end{abstract}

\maketitle




{\it Introduction.}---Understanding whether and how a quantum many-body system equilibrates is a question that permeates many areas of physics. One way to probe equilibration is through entanglement, which is increasingly being used to characterize quantum matter. We can also view the process of equilibration as a useful dynamical setting to study the generation of entanglement between subsystems, a question of much interest in quantum information theory.

In holographic duality, equilibration from a generic initial many-body state maps to black hole formation from a gravitational collapse, and questions related to equilibration become intimately connected to issues in black hole physics. This connection, on one hand, brings in powerful gravity techniques to deal with non-equilibrium systems, and on the other, gives new perspectives on black holes. 


In this paper, we consider the evolution of entanglement entropy after a global quench in a strongly coupled {\it gapless} system with a gravity dual, extending earlier results in~\cite{Calabrese:2005in,Hubeny:2007xt,AbajoArrastia:2010yt,Albash:2010mv,Balasubramanian:2010ce,Aparicio:2011zy,Galante:2012pv,Caceres:2012em,Baron,Arefeva:2013wma}. We find that the entanglement entropy exhibits a variety 
of scaling behavior which lead to a strikingly simple geometric picture for entanglement growth.

{\it Set-up.}---At $\ft =0$ we turn on external sources for an interval $ \de \ft$ in
a $d$-dimensional  boundary system, creating a {\it spatially homogeneous and isotropic} excited state with nonzero energy density, which subsequently equilibrates. 
We work in the {\it quench} limit, taking the sourcing interval $\de \ft$ to zero. On the gravity side, such a quench process is described by an infinitesimally thin shell of matter which collapses to form a black hole, and can be modeled by an AdS-Vaidya metric 
of the form 
\be \label{vaidya}
ds^2 = {L^2 \ov z^2} \le(- f (v,z) dv^2 - 2 dvdz + d \vec x^2 \ri) 
\ee
with $f (v,z) = 1-  \th (v) g(z)$. Our results will not depend on the content and configuration of matter fields making up the shell (which are determined by the boundary sourcing process); in the classical gravity regime we are working with, which translates to the large $N$ and strongly coupled limit of the boundary theory, the entanglement entropy is sensitive only to the metric of the collapsing geometry.  

For $v< 0$, the metric~\eqref{vaidya} is given by that of pure AdS describing the vacuum state before the quench, and for $v > 0$ it is given by that of a black hole, 
\be \label{BHre1}
ds^2 = {L^2 \ov z^2} \le(- h (z) dv^2 - 2  dvdz + d \vec x^2 \ri) \ ,
\ee
where the function $h(z) \equiv 1 - g(z)$ can be viewed as parameterizing different final equilibrium states. Seeking ``universal" behavior among different equilibrium states, we  consider a general $h(z)$ satisfying the following properties: (i) it has a simple zero at $z= z_h$, i.e. the event horizon of~\eqref{BHre1}, (ii) near the boundary $z=0$, $h(z) =1 - M z^d +\cdots$ for some $M$, (iii) for $z < z_h$, it is positive and monotonically decreasing as a function of $z$ as required by the IR/UV connection, (iv) the metric~\eqref{BHre1} satisfies the null energy condition. We assume that any such $h(z)$ can be realized by a suitable arrangement of matter fields. Representative examples of~\eqref{BHre1} include the AdS-Schwarzschild black hole with $g(z) = M z^d$, and the AdS Reissner-Nordstrom (RN) black hole with $g(z) = M z^d - Q^2 z^{2 d-2}$ describing an equilibrium state with nonzero charge density proportional to $Q$. 
The temperature, energy, and entropy density of the equilibrium state corresponding to \eqref{BHre1} are given by $T = {|h'(z_h)| \ov 4 \pi}$, $\sE = {L^{d-1} \ov 8 \pi G_N} {d-1 \ov 2} M$, and $s_{\rm eq} = {L^{d-1} \ov z_h^{d-1}} {1 \ov 4 G_N}$, where $G_N$ is Newton's constant in the bulk.

Now let us consider a spatial region in the boundary theory bounded by a {\it smooth} surface $\Sig$. The entanglement entropy of this region $S_{\Sig} (\ft)$ can be obtained using $S_{\Sig} (\ft) = {\sA_\Sig \ov 4 G_N}$, where 
$\sA_\Sig$ is the area of a $(d-1)$-dimensional bulk extremal surface ending at $\Sig$ on the boundary~\cite{Ryu:2006bv,Hubeny:2007xt}. We define the size $R$ of $\Sig$ to be the height of its future domain of dependence.\footnote{We thank V.~Hubeny for this suggestion.}         

Denote by $\De S_\Sig (\ft)$ the difference of the entanglement entropy with that of the vacuum.
After the quench, $\De S_{\Sig} (\ft)$ starts evolving at $\ft =0$ from $0$ to
the equilibrium value $\Delta S^{\rm (eq)}_\Sig = s_{\rm eq} V_\Sig$ (for sufficiently large $R$), where $V_\Sig$ is the volume of the region bounded by $\Sig$. With the sourcing interval $\de \ft = 0$, there exists a sharp saturation time $\ft_s$ when $S_{\Sig} (\ft)$ saturates at $S^{\rm (eq)}_\Sig$ and remains constant afterwards~\cite{Calabrese:2005in,AbajoArrastia:2010yt,Albash:2010mv,Balasubramanian:2010ce}. 

To describe our results we first introduce a ``local equilibrium scale'' $\ell_{\rm eq}$, 
which can be
defined as the time scale when the system has ceased production of thermodynamic entropy. For an equilibration process described by~\eqref{vaidya}, we identify it as
$\eql \sim z_h  \sim  ({1 / s_{\rm eq}})^{1 \ov d-1}$. 
For $h(z)$ given by a Schwarzschild black hole, $\ell_{\rm eq} \sim {1 /T}$, while for a RN black hole $\eql$ can be much smaller than $1/T$ when $Q$ is large. 
We emphasize that at times of order $\eql$, while local thermodynamics already applies at scales comparable to or smaller than $\eql$,  the entanglement entropy $S_{\Sig} (\ft)$ (and other nonlocal observables defined over large distances) remain far from their equilibrium values, for $R \gg \eql$. 

We are interested in probing quantum entanglement at macroscopic scales, and take $R \gg \eql$. On the gravity side, there is an important geometric distinction between extremal surfaces for $R \lesssim \eql$ and for $R \gg \eql$. For $R \lesssim \eql$, the evolution lasts for a time of order $\ft \lesssim \eql$ and the extremal surface stays outside the horizon during the entire evolution. For $R \gg \eql$, the evolution is controlled by the geometry around and inside the horizon 
for $\ft \gtrsim \eql$. Turning this around, we  
can use entanglement entropy (and other nonlocal observables such as Wilson loops and correlation functions) to probe the geometry behind the horizon of a collapsing black hole. This is in contrast to the static case of an eternal black hole where  
extremal surfaces always stay outside the horizon~\cite{Hubeny:2012ry}.

{\it Results.}---To find $S_{\Sig} (\ft)$, one proceeds to solve the equations for an extremal surface in~\eqref{vaidya} with given boundary conditions and evaluate its area. By identifying various geometric regimes for the bulk extremal surface, we can extract the analytic behavior of $S_{\Sig} (\ft)$ during corresponding stages of time evolution. An important observation is that of the existence of a family of ``critical extremal surfaces'' which lie behind the horizon and separate extremal surfaces that reach the boundary from those which fall into the black hole singularity. In particular, for $R, \ft \gg \eql$, the leading behavior of the entanglement entropy can be obtained from the geometry of such critical extremal surfaces. Here we discuss our main results and their physical implications, leaving a detailed technical exposition to elsewhere~\cite{josenew}. See Appendix A for a brief 
discussion of geometric regimes corresponding to the stages of evolution discussed below.  

{\bf 1. Pre-local-equilibration growth:} For $\ft \ll \eql$, the entanglement entropy 
 grows as 
\be \label{quda}
\De S_\Sig (\ft) = {\pi \ov d-1} \sE A_\Sig \ft^2 + \cdots
\ee
where  $\sE$ is the energy density and $A_{\Sig}$ is the area of $\Sig$. This result is independent of the shape of $\Sig$, the spacetime dimension $d$, and the specific form of $h(z)$.\footnote{For $d=2$ with $h(z)$ give by that of a BTZ black hole, the quadratic time dependence was  also obtained in~\cite{Hubeny:2013hz}, and are implicit in~\cite{Balasubramanian:2010ce,Aparicio:2011zy}.} 
 
 {\bf 2.  Post-local-equilibration linear growth:} For $R \gg \ft \gg \eql$, we find a universal linear growth   
\be \label{line}
\De S_\Sig (\ft) =\Vee s_{\rm eq}  A_\Sig \ft + \cdots
\ee
where $\Vee$ is a dimensionless number which is {\it independent} of the shape of $\Sig$, but does depend on the final equilibrium state. It is given by  
\be \label{gene}
\Vee =  ({z_h / z_m})^{d-1} \sqrt{- h(z_m)} 
\ee
where $z_m$ is a minimum of ${h(z) \ov z^{2 (d-1)}}$ 
and lies inside the horizon. See Fig.~\ref{fig:critical} of Appendix A for an illustration of extremal surfaces in this regime. 
For~\eqref{BHre1} given by a Schwarzschild black hole, 
\be \label{schwv}
\Vee^{(\rm S)} = {(\eta -1)^{\ha (\eta -1)} \ov \eta^{\ha \eta}}  \ , \quad \eta \equiv  {2 (d-1) \ov d} 
\ee
while for a Reissner-Norstrom black hole,
\be \label{stvrn}
\Vee^{\rm (RN)} =  \sqrt{1 \ov \eta -1} \le(\le(1- { u \ov \eta}  \ri)^{\eta} - (1-u) \ri)^\ha 
\ee
where $u \equiv {4 \pi z_h T \ov d}$ is $1$ for $Q=0$, and decreases monotonically 
to zero as $Q$ is increased to $\infty$ (at fixed $T$). Note that $\Vee^{(S)} = 1$ for $d=2$ and monotonically decreases with $d$, while  $\Vee^{\rm (RN)}$ monotonically increases with $u$ -- turning on a nonzero charge density slows down the evolution.~\eqref{line} generalizes previous observations of linear growth for $d=2$~\cite{Calabrese:2005in,AbajoArrastia:2010yt}.  With a different set-up in the bulk, the linear growth~\eqref{line} as well as~\eqref{gene}--\eqref{schwv} 
were obtained recently in~\cite{Hartman:2013qma}. 

 {\bf 3. Saturation:}  The evolution beyond the linear regime depends on the shape, the spacetime dimension $d$, and may also depend on the final equilibrium state.  We  focus on the most symmetric shapes, with $\Sig$ given by a sphere or strip and $R$ the radius of the sphere or half-width of the strip. We will denote the entanglement entropy for a sphere as $S (R, \ft)$, and its equilibrium value as $S_{\rm eq} (R)$.
For a strip in $d \geq 3$, the linear behavior~\eqref{line} persists all the way to saturation with saturation time given by
\be \label{stripS}
\ft_s^{(\rm strip)} = {S_{\rm strip}^{\rm (eq)} \ov \Vee s_{\rm eq} A_{\rm strip}} + \cdots = {R \ov v_E}  + O(R^0) \ ,
\ee 
at which the bulk extremal surface jumps 
discontinuously. This is analogous to a first-order phase transition with the first derivative of $S_\Sig (\ft)$ being discontinuous at $\ft_s$.\footnote{With an RN black hole, at fixed $Q$ and sufficiently large $R$, the linear regime does not appear to extend all the way to saturation.} Similar behavior also happens for a sphere in $d=3$ when the final state is given by a RN black hole with sufficiently large $Q$.\footnote{The discontinuous 
saturation for a strip and for a sphere in $d=3$ for a RN black hole with a sufficiently large $Q$ was first observed in~\cite{Albash:2010mv}. Also see~\cite{Balasubramanian:2010ce}.}

For a sphere in $d \geq 4$ and any $h(z)$, the approach to saturation resembles that of 
a continuous phase transition and can be characterized by a nontrivial scaling exponent 
\be \label{svew}
 S (R,\ft)- S^{\rm (eq)} (R) \propto - (\ft_s - \ft)^\ga\ , \qquad \ga = {d+1 \ov 2}
 \ee
for $\ft_s  - \ft \ll \eql$. 
The same exponent also applies in $d=2$ for a BTZ black hole as was recently found in~\cite{Hubeny:2013hz}.
In $d=3$, for a continuous saturation, 
we find $S (\ft, R)- S_{\rm eq} (R) \propto  (\ft_s - \ft)^2 \log  (\ft_s - \ft)$, with the logarithmic scaling barely avoiding the ``mean-field" exponent $\gamma=2$. Meanwhile, the saturation time for a sphere, again in cases where the saturation is continuous, is given by
\be \label{fjske}
\ft_s (R)= {1 \ov c_E } R - {d-2 \ov 4 \pi T} \log R + O(R^0)
\ee
where $c_E$ is the dimensionless number 
\be \label{uber}
c_{E} = \sqrt{z_h |h'(z_h)| \ov 2(d-1)} = \sqrt{2 \pi z_h T \ov d-1} \ ,
\ee
which for Schwarzschild and RN black holes becomes 
\be 
c_E^{(\rm S)} = 1/\sqrt{\eta}\ ,   \quad c_E^{(\rm RN)} = \sqrt{u/ \eta} \leq c_E^{(S)}  \ .
\ee
Note that for $d=2$, $c_E =1$ and the logarithmic term in~\eqref{fjske} 
disappears, which gives the results in~\cite{Calabrese:2005in,AbajoArrastia:2010yt,Balasubramanian:2010ce}.

{\bf 4. Late-time memory loss}: For a sphere, there is an additional scaling regime for $\ft_s \gg \ft_s - \ft \gg \eql$ in which $S (R, \ft)$ only depends on the difference $\ft_s (R) - \ft$, and not on $\ft$ and $R$ separately.\footnote{The expression below may not apply in $d=3$, where we do not yet have a clean result.} There
\be \label{evem}
S (R, \ft) - S_{\rm eq} (R) = - s_{\rm eq} \lam \le(\ft_s (R) - \ft \ri) 
\ee
where $\lam$ is  some  function that depends on $h(z)$ and which we have only determined for $d=2$ with the BTZ black hole,  
\be \label{evem1}
\lam (y) =  \le[y + {1 \ov 2 \pi T} \log \le( \sin \chi^{-1} (2 \pi T y)\ri) \ri], 
\ee
with $ y = R -\ft$ and $\chi (\phi) =  \le(\cot {\phi \ov 2} -1 \ri)  + \log \tan {\phi \ov 2}$. Note $\lam (y)$ interpolates between the linear behavior~\eqref{line} for large $2 \pi T y$ 
and the critical behavior $y^{3 \ov 2}$ near saturation as $2 \pi T y \to 0$. Bulk extremal surfaces in this late-time regime trace the horizon--see Fig.~\ref{fig:critical} in Appendix A for an illustration. 
For $d=2$, the equivalent of~\eqref{evem} was first observed in~\cite{AbajoArrastia:2010yt,Aparicio:2011zy}.

\begin{figure}[!h]
\begin{center}
\includegraphics[scale=0.3]{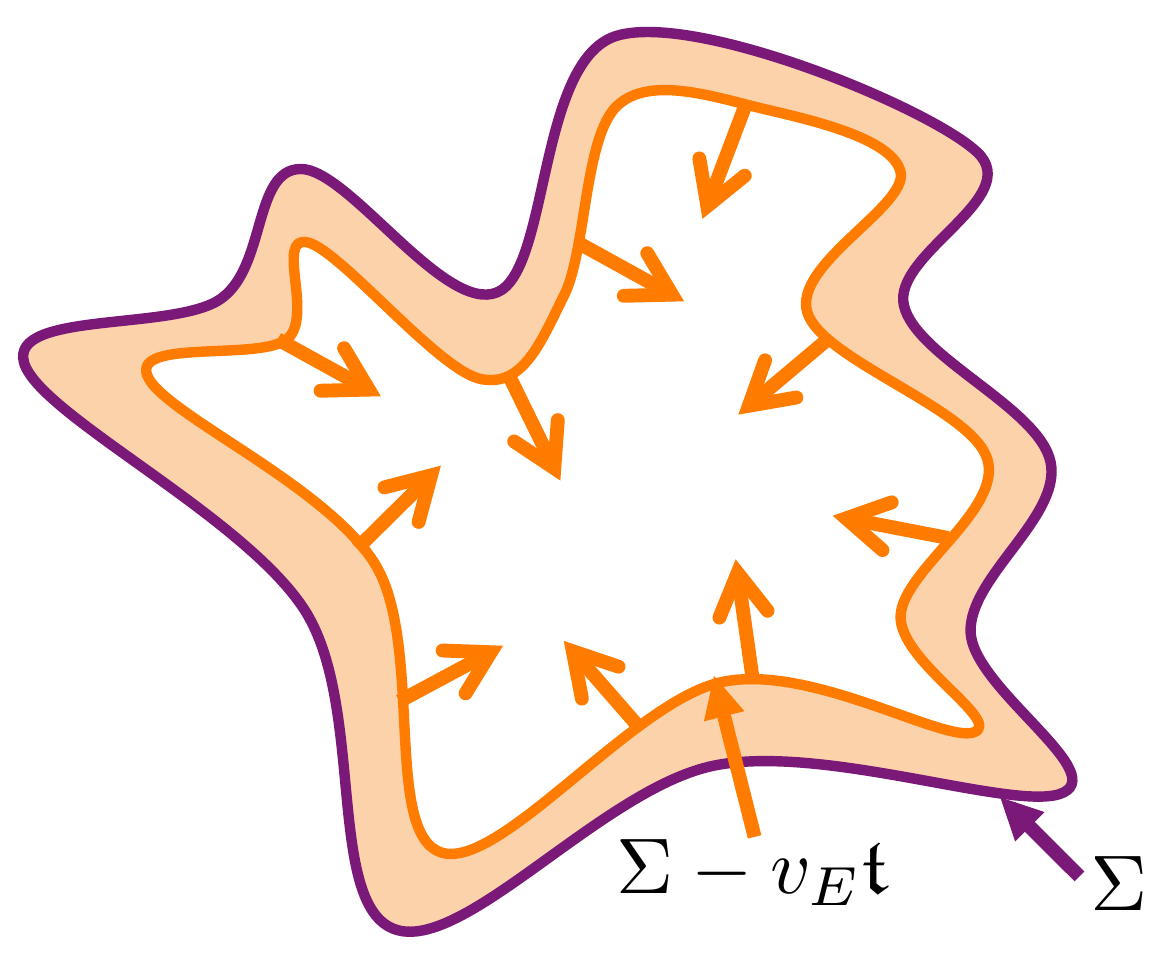} \quad
\includegraphics[scale=0.4]{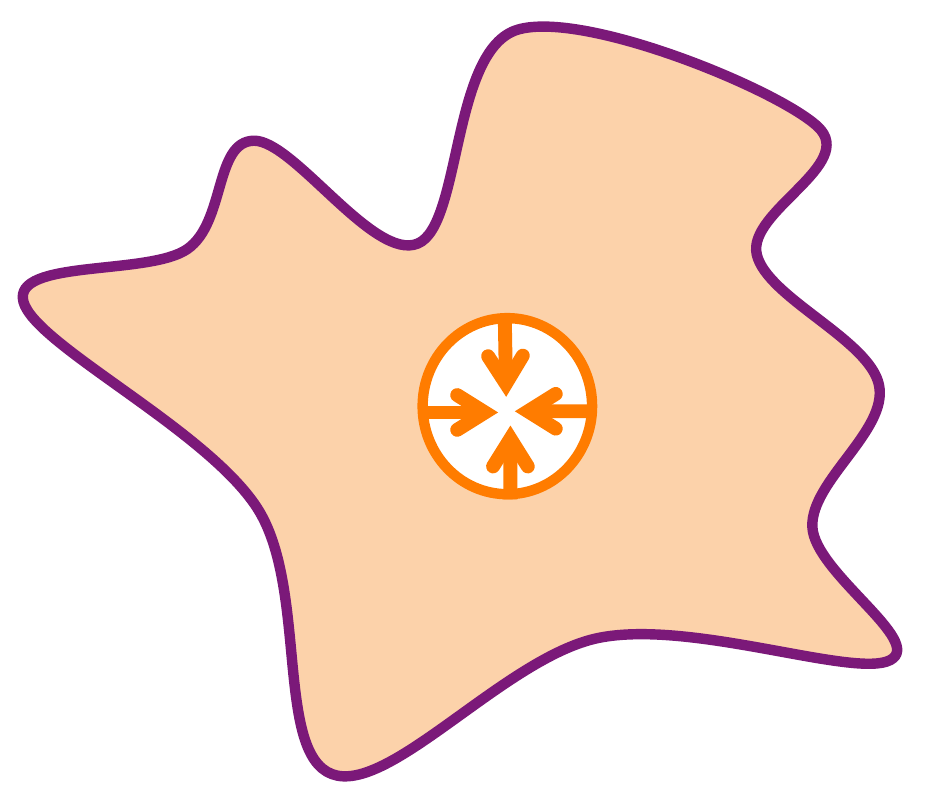}
\end{center}
\caption{{\it Left}: The growth in entanglement entropy can be visualized as occuring via an ``entanglement tsunami" with a sharp wave-front carrying entanglement inward from $\Sig$. {\it Right}: Late-time memory loss--for a wide class of compact $\Sig$, in the limit of large size, at late times the wave front may approach 
that of a spherical $\Sig$.
 }\label{fig:eevelo}
\end{figure}

{\it Physical interpretation.}---Equation~\eqref{line} can be rewritten
\be \label{eirp}
\De S_{\Sig} (\ft) = s_{\rm eq} \le(V_{\Sig} - V_{\Sig - \Vee \ft} \ri) + \cdots 
\ee
where $V_{\Sig - \Vee \ft}$ denotes the volume of the region bounded by a surface which is obtained from $\Sig$ by moving every point inward by a distance $\Vee \ft$ (see Fig.~\ref{fig:eevelo}). 
This suggests a simple geometric picture of entanglement growth: there is a wave with a sharp wave-front propagating inward from $\Sigma$, and the region that has been covered by the wave is entangled with the region outside $\Sig$, while the region yet to be covered (i.e. orange region in Fig.~1) is not so entangled. We dub this wave an ``entanglement tsuanmi". In this picture, saturation occurs when the tsunami covers the full region.

The tsunami picture suggests that the evolution of entanglement is {\it local}. This is natural as the time evolution in our system is generated by a {\it local} Hamiltonian which couples directly only to the degrees of freedom near $\Sig$, so the entanglement has to build up from $\Sig$. When $R$ is large, the curvature of $\Sig$ is negligible at early times, which explains the area law and shape-independence of~\eqref{quda} and~\eqref{line}, i.e. different parts of the tsunami do not interact with one another. As the tsunami advances inward, curvature effects become important, and  the propagation should become nonlinear due to ``interactions'' among different parts of the tsunami, resulting in shape-dependent saturation.

For a strip in $d \geq 3$, equation~\eqref{stripS} suggests tsunamis from two boundaries propagate freely until they meet each other at which time saturation happens discontinuously. 

For a sphere in $d \geq 3$,  $1 > c_E > \Vee$, where the latter inequality may be understood geometrically from the fact that the volume of an annulus region of unit width becomes smaller as the tsunami advances inward. The former inequality, along with the presence of the logarithm in~\eqref{fjske} and the nontrivial exponent $\ga$ in~\eqref{svew}, may be considered a consequence of interactions among different parts of the tsunami. In contrast, in $d=2$ where $\Sig$ consists of two points, $c_E = \Vee = 1$. 

Finally, given $S_{\rm eq} (R) = V_{\Sig} s_{\rm eq}$, one can interpret $\lam$ in~\eqref{evem}
as the volume which has not yet been entangled, and ~\eqref{evem} as implying that that volume only depends on the difference $\ft_s - \ft$ and not on $R$ and $\ft$ separately. In other words, at late times of evolution, the size $R$ has been ``forgotten''.  We emphasize that with \eqref{evem} valid for $\ft_s \gg \ft_s - \ft \gg \eql$, such memory loss can happen long before saturation. It is tempting to speculate that for a generic surface $\Sig$, in the limit of large $R$, memory of {\it both} the size and shape of $\Sig$ could be lost during late times of evolution. This would also imply that the critical behavior~\eqref{svew} for a sphere may in fact apply to a wider class of compact surfaces near saturation. In other words, for such surfaces, at late times the wave front of the tsunami may approach an ``IR fixed point'' provided by the sphere. See Fig.~\ref{fig:eevelo} for an illustration.

The linear growth regime~\eqref{line} sets in only after local equilibration has been achieved. This explains the appearance of the equilibrium entropy density $s_{\rm eq}$ in the prefactor. 
In contrast, the pre-local-equilibration quadratic growth~\eqref{quda} is proportional to the energy density $\sE$. Indeed, at very early times before the system has equilibrated 
locally, the only macroscopic data characterizing the initial state is the energy density. 
Conversely, if we stipulate that before local equilibration $S_{\Sig} (\ft)$ should be proportional to $A_{\Sig}$ and $\sE$, the quadratic time dependence in$~\eqref{quda}$ follows from dimensional analysis. Similarly, if we require that after local equilibration, $S_{\Sig} (\ft)$ is proportional to $A_{\Sig}$ and $s_{\rm eq}$, linearity in time follows. 

In~\cite{Calabrese:2005in}, a model of entanglement growth from 
free-streaming ``quasiparticles'' was proposed which gave a nice explanation for the linear growth and saturation of entanglement entropy in $d=2$. In particular,
$\Vee= c_E = 1$ followed from quasiparticles propagating at the speed of light. 
 In a companion paper with M\'ark Mezei~\cite{speed}, we generalize the free-streaming model to higher dimensions and again find that at early times there is linear growth as in~\eqref{line} with $s_{\rm eq}$ interpreted as giving a measure for quasiparticle density.\footnote{The quasiparticle model can also be generalized to capture the pre-local-equilibration quadratic growth~\eqref{quda}, if one takes into account that  during local equilibration the quasiparticle density gradually builds up~\cite{speed}.}
Since the quasiparticles can travel in different directions in $d\geq 3$, although their individual speed is set to $1$, the speed of the entanglement tsunami is smaller than $1$, and is given by~\cite{speed} 
\be \label{vstrem}
v^{\rm (streaming)}_E = { \Ga({d-1 \ov 2}) \ov \sqrt{\pi}  \Ga ({d \ov 2}) } <  \Vee^{(S)} < 1 \ .
\ee
We note that this speed is in fact smaller than even the Schwarzschild value~\eqref{schwv}--in strongly coupled systems, the propagation of entanglement entropy is faster than that from free-streaming particles moving at the speed of light! Recall that a hallmark of a strongly coupled system with a gravity dual is the {\it absence} of a quasiparticle description. Thus while the quasiparticle model appears to capture linear growth, it is likely missing some important elements present in a holographic system, e.g. multi-body entanglement.
Also, note that in the quasiparticle model, $\ft_s = R$ for a sphere, implying faster saturation that in~\eqref{fjske}, as $c_E < 1$ for $d \geq 3$.

{\it Generality.}---Since the qualitative picture of an entanglement tsunami followed from evolution under a local Hamiltonian, we expect it to apply 
to more general equilibration processes for which the initial state does not have to be homogeneous or isotropic, and even to systems without translation-invariance. With a nonzero sourcing interval $\de \ft$, the 
wave front will develop a finite spread, but the picture of an entanglement wave that propagates
may still apply as long as $\de \ft$ is much smaller than the 
size of the region one is exploring. If $\de \ft$ is comparable to or larger than $\eql$, the 
pre-local-equilibration and saturation regimes can no longer be sharply defined, but post-local-equilibration linear growth should still exist, as could late-time memory loss. An important feature of the linear growth~\eqref{line} is that the tsunami speed $\Vee$ {\it characterizes properties of the equilibrium state}, as it is solely determined by the metric of the black hole. This again highlights the local nature of entanglement propagation. At corresponding times, locally, the system has already achieved equilibrium, although for large regions non-local observables such as entanglement entropy remain far from their equilibrium values. Thus  $\Vee$ should be independent of the nature of the initial state, including whether it was isotropic or homogeneous.
Finally, that the early growth~\eqref{quda} is proportional to the energy density is consistent with other recent studies of the entanglement entropy of excited states~\cite{Bhattacharya:2012mi,blancoetal,Allahbakhshi:2013rda,Wong:2013gua}. 

{\it Maximal entanglement rate?}---To be able to compare the growth of entanglement entropy among different systems, we introduce a dimensionless rate of growth
\be \label{groR}
\fR_{\Sig} (\ft) \equiv {1 \ov s_{\rm eq} A_{\Sig}} {d S_{\Sig} \ov d\ft} \ .
\ee
In the linear regime, $\fR_{\Sig}$ is a constant given by $\Vee$, while in the pre-local-equilibration 
regime $\ft \ll \eql$ in which~\eqref{quda}, 
\be 
\fR_{\Sig} (\ft)  = {2 \pi \ov d-1} {\sE \ft \ov s_{\rm eq}} 
\ee 
grows linearly with time.

 In a relativistic system, $\Vee$, and more generally $\fR_{\Sig} (\ft)$, 
should be constrained by causality, although relating them directly to the speed of light appears difficult except 
in the quasiparticle-type model mentioned earlier. We have examined $\Vee$  for known black hole solutions and also various $h(z)$ satisfying the properties listed below~\eqref{BHre1}, and find support that 
\be \label{emrp}
\Vee  \leq \Vee^{(\rm S)} = {(\eta -1)^{\ha (\eta -1)} \ov \eta^{\ha \eta}} 
= \bca 1 &  d=2 \cr
          {\sqrt{3} \ov 2^{4 \ov 3}} = 0.687 & d=3 \cr
          {\sqrt{2} \ov 3^{3 \ov 4}} = 0.620 & d=4 \cr
          \ha & d = \infty 
          \eca \ .
\ee
See Appendix B for some explicit examples. That~\eqref{BHre1} satisfies the null energy condition appears to be important for the validity of the above inequality.\footnote{An alternative way to eliminate some of the examples which violate the above inequalities is to require the positivity of relative entropy between the excited state under consideration and the vacuum~\cite{blancoetal}.} There are reasons to suspect that the Schwarzschild value~\eqref{schwv} may indeed be special. The gravity limit corresponds to the infinite coupling limit of the gapless boundary Hamiltonian, in which the generation of entanglement should be most efficient.  From the bulk perspective,
it is natural to expect that turning on additional matter fields (satisfying the null energy condition) will slow down thermalization. From the boundary perspective, this is consistent with one's expectation that when there are conserved quantities such as charge density, the equilibration process should become less efficient. Given~\eqref{vstrem}, it is tempting to conjecture that~\eqref{emrp} applies to {\it all} relativistic systems for which linear growth regime exist. 

Since the bound~\eqref{emrp} is saturated for the Schwarzschild black hole in Einstein gravity, higher derivative corrections to the Einstein action could be a particularly dangerous source of violation. A general holographic prescription for computing entanglement entropy in higher derivative gravities is not yet available except in the case of Gauss-Bonnet gravity in $d=4$, for which a proposal was made in~\cite{deBoer:2011wk,Hung:2011xb}. Using their proposal, we find 
\be
v_E = {\sqrt{2} \ov 3^{3 \ov 4}} - {3^{1 \ov 4} \ov \sqrt{2}} \lam+ O(\lam^2)  
\ee
where $\lam$ is the Gauss-Bonnet coupling. While in principle $\lam$ can take both signs, in all known 
examples $\lam$ appears to be positive~\cite{Buchel:2008vz}. We should also note that in all known examples where the Gauss-Bonnet term arises, there are probe branes and orientifolds which back-react on the metric and give rise to additional contributions at the same (or a more dominant) order. Thus it appears we cannot draw a conclusion at this point. 

Turning to $\fR_\Sig$, in explicit examples we find that after local equilibration (i.e. after the linear growth regime has set in), it monotonically decreases with time. This appears natural from the tsunami picture, 
as after the linear regime ``interactions'' among different parts of the tsunami are likely to slow it down. So it is tempting to speculate that {\it after local equilibration}
\be \label{ineq2}
\fR_\Sig (\ft) \leq \Vee^{(\rm S)} \ .
\ee
Before local equilibration, however, the behavior of $\fR_\Sig$ is sensitive to the initial state, and  in particular for a RN black hole with $\Sig$ a sphere or strip, we find $\fR_\Sig$ can exceed $\Vee^{(\rm S)}$ 
near $\eql$ (see Fig.~\ref{fig:frbe} in Appendix B). Also, for a highly anisotropic initial state, $\fR_\Sig$ could for a certain period of time resemble that of a $(1+1)$-dimensional system for which it can reach $1$.
\footnote{We thank T.~Hartman for a discussion on this point.} Thus we speculate that before local equilibration,
\be \label{eorp}
\fR_\Sig (\ft) \leq 1 \ .
\ee

The inequalities \eqref{emrp}, \eqref{ineq2} and~\eqref{eorp} are reminiscent of the small incremental entangling conjecture~\cite{bravyi} for ancilla-assisted entanglement rates in a {\it spin} system, which was recently proved in~\cite{franketal}. The conjecture states that ${dS \ov dt} \leq c ||H|| \log D$ where $S$ is the entanglement entropy between subsystems $aA$ and $bB$, $||H||$ is the norm of the Hamiltonian $H$ that generates entanglement between $A$ and $B$ ($a$, $b$ are ancillas), $D = {\rm min} (D_A, D_B)$ where $D_A$ is the dimension of the Hilbert space of $A$, and $c$ is a constant independent of $D$.
In our case, the Hamiltonian is local and thus couples directly only to the degrees of freedom near $\Sig$--the analogue of $\log D$ is proportional to $A_\Sig$, and the entropy density $s_{\rm eq}$ in~\eqref{groR} can be seen as giving a measure of the density of excited degrees of freedom. 

Finally, it would be interesting to formulate an effective theory for the propagation of an entanglement tsunami in a general equilibration process. This would be similar in spirit to recent efforts in~\cite{Nozaki:2013vta} to derive equations of motion for entanglement entropy from Einstein equations.

{\it Note Added:} after the appearance of this paper we received~\cite{Li:2013sia,Shenker:2013pqa} which have some overlap with our study.

\vspace{0.2in}  
{\bf Acknowledgements}---
We thank  M\'ark Mezei for many discussions, and also A.~Adams, H.~Casini, T.~Faulkner, A.~Harrow, T.~Hartman, M.~Headrick, V.~Hubeny, L.~Huijse, S.-S. Lee, J.~Maldacena, R.~Myers, S.~Pufu, M.~Rangamani, 
O.~Saremi, D.~T.~Son, J.~Sonner, B.~Swingle, T.~Takayanagi, E.~Tonni, and J.~Zaanen. Work was supported in part by funds provided by the U.S. Department of Energy
(D.O.E.) under cooperative research agreement DE-FG0205ER41360 and Simons Foundation.


    

\appendix 

\section{Appendix A: Geometric regimes for bulk extremal surfaces}

Here we briefly describe geometric features of bulk extremal surfaces corresponding to various stages of evolution described in the main text. Note we take $R$, the size of $\Sig$, to be much greater than $\eql$. 

For a given $\Sig$, at time $\ft$, the corresponding bulk extremal surface which we denote as $\Ga_\Sig (\ft)$ has a tip corresponding to its furthest point in the bulk. 
We denote the location of this tip as $(z_t (\ft), v_t (\ft))$.  There can be multiple extremal surfaces for a given $(\Sig, \ft)$ and we should choose the one with smallest area. 
When $\Sig$ is a sphere or strip, specifying $(z_t, v_t)$ completely fixes $\Ga_{\Sig}$, although the relations between $(R, \ft)$ and $(z_t, v_t)$ are in general complicated and require solving the full equations of motion for $\Ga_{\Sig}$. 

\begin{figure}[!h]
\begin{center}
\includegraphics[scale=0.4]{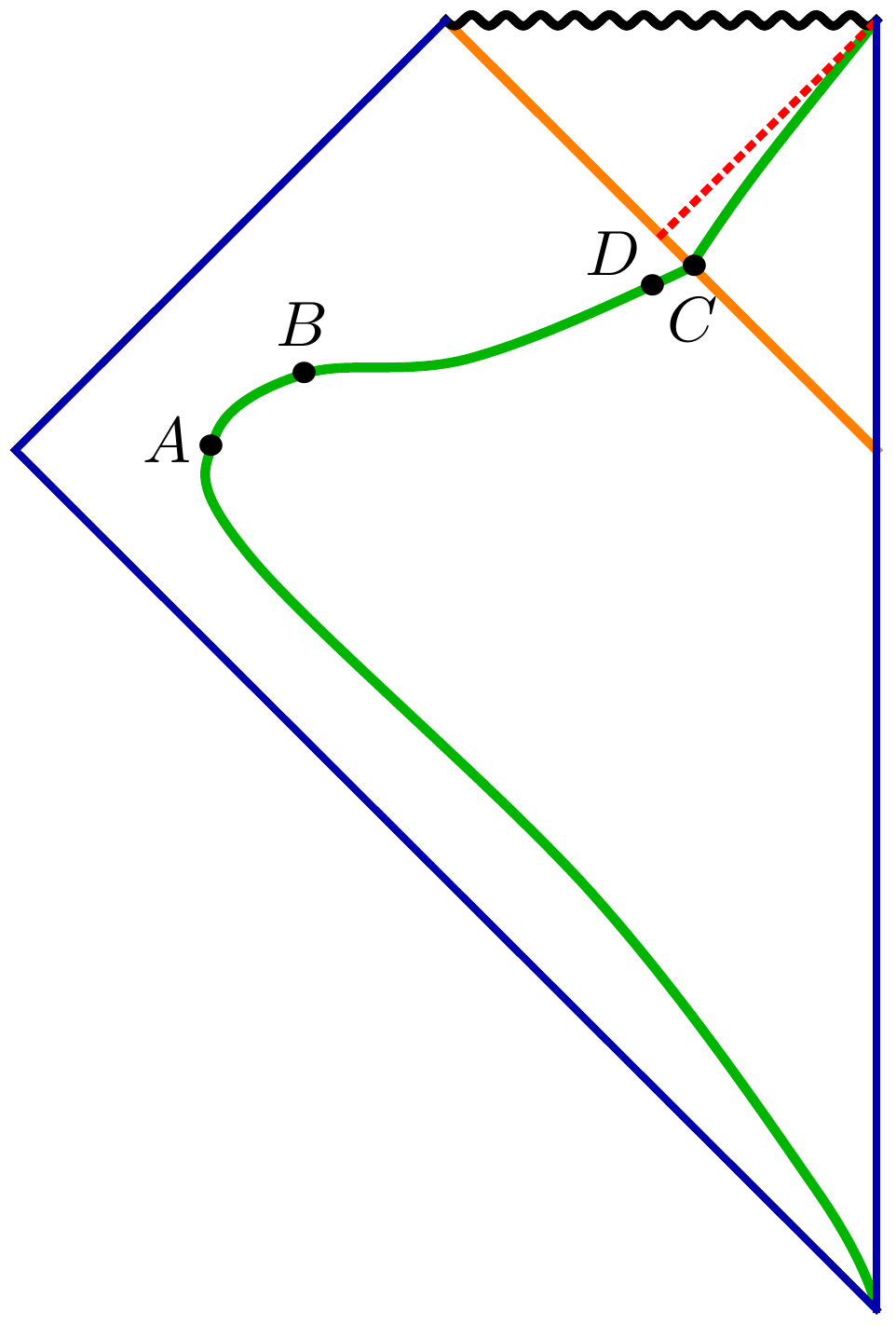}   \includegraphics[scale=0.4]{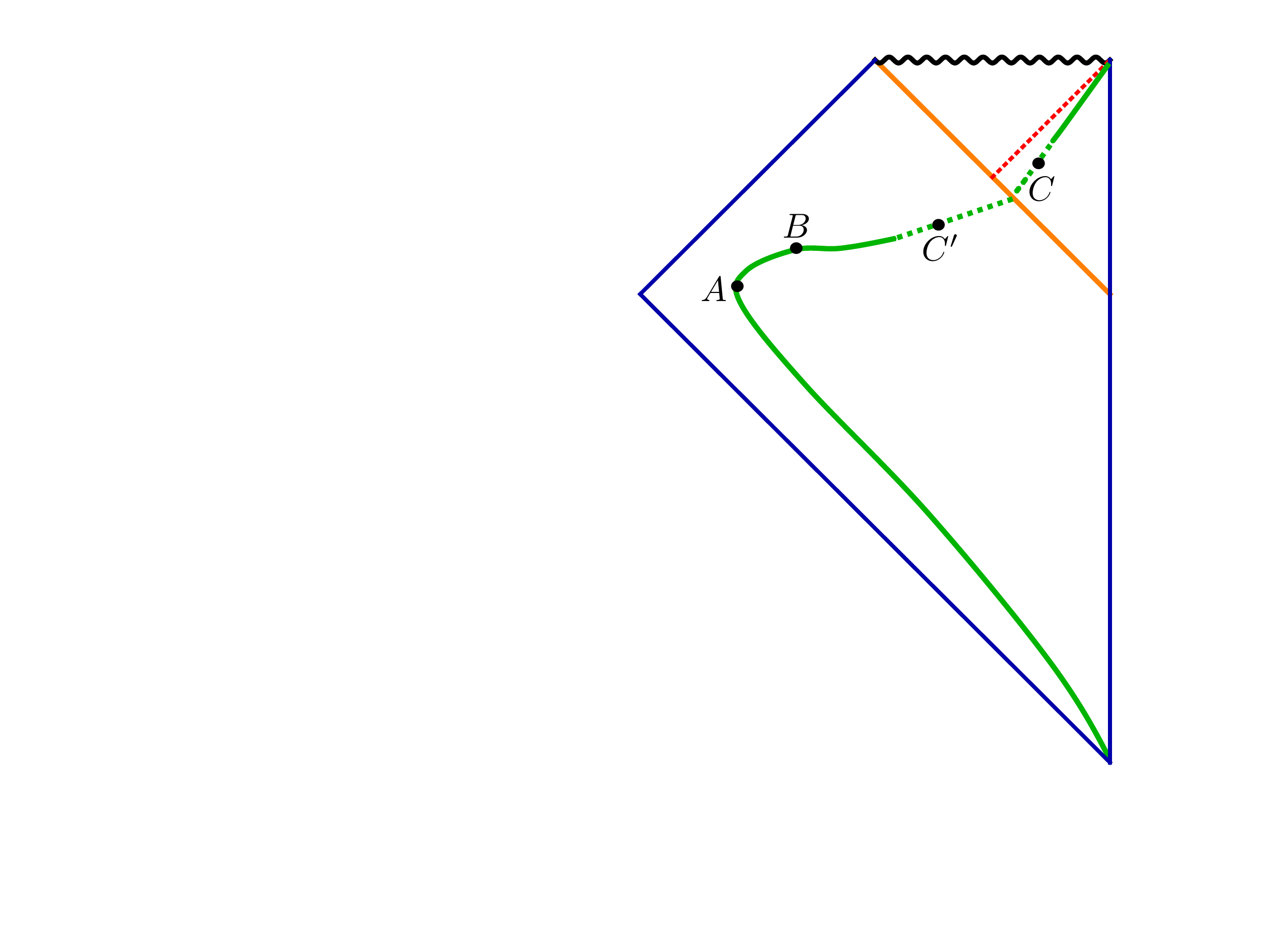}
\end{center}
\caption{Cartoons of the curve $(z_t (\ft), v_t (\ft))$ at fixed $\Sig$ in a Penrose diagram for continuous (left) and discontinuous (right) saturation. The solid orange line denotes the in-falling matter shell and the red dashed line is the black hole horizon. Time progresses from bottom to top and boundary spatial directions are suppressed.
{\it Left}: For continuous saturation the whole curve is single-valued throughout, and saturation happens at point $C$. {\it Right}: Discontinuous saturation happens via a jump between two branches of the curve, from $C'$ to $C$. On the dashed portion of the curve, different points $(z_t, v_t)$ can correspond to the same $\ft$.
}
\label{fig:curves}
\end{figure}

\begin{figure}[!h]
\begin{center}
\includegraphics[scale=0.5]{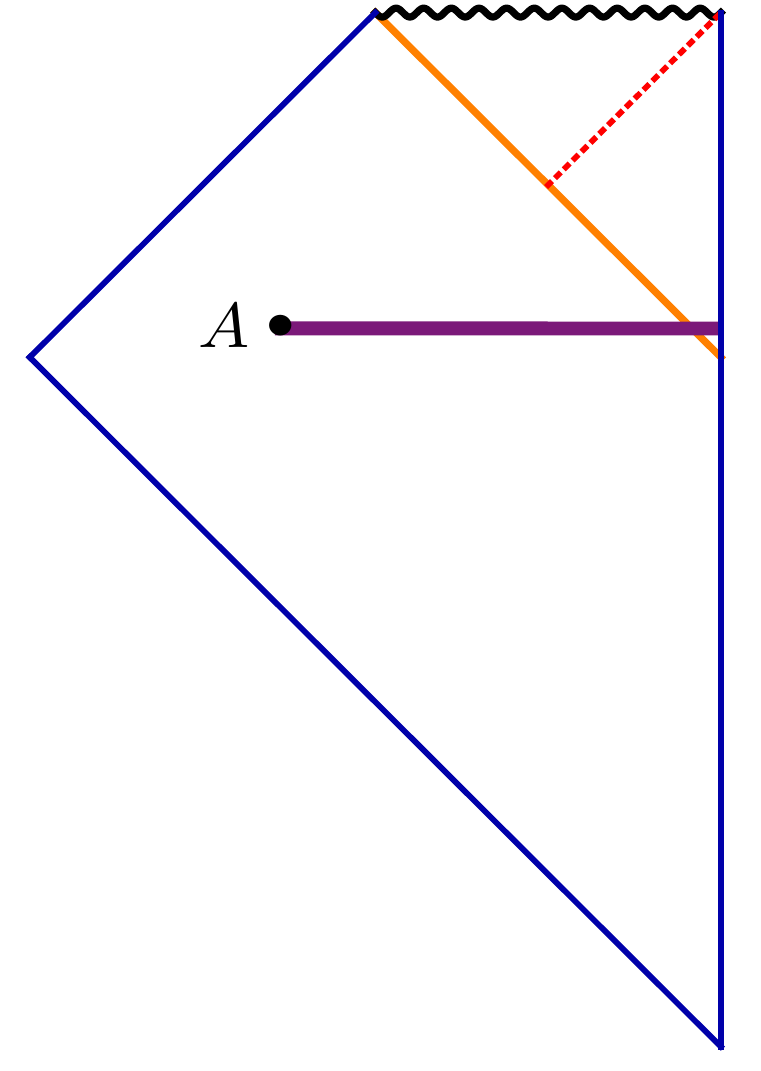} \;
\includegraphics[scale=0.5]{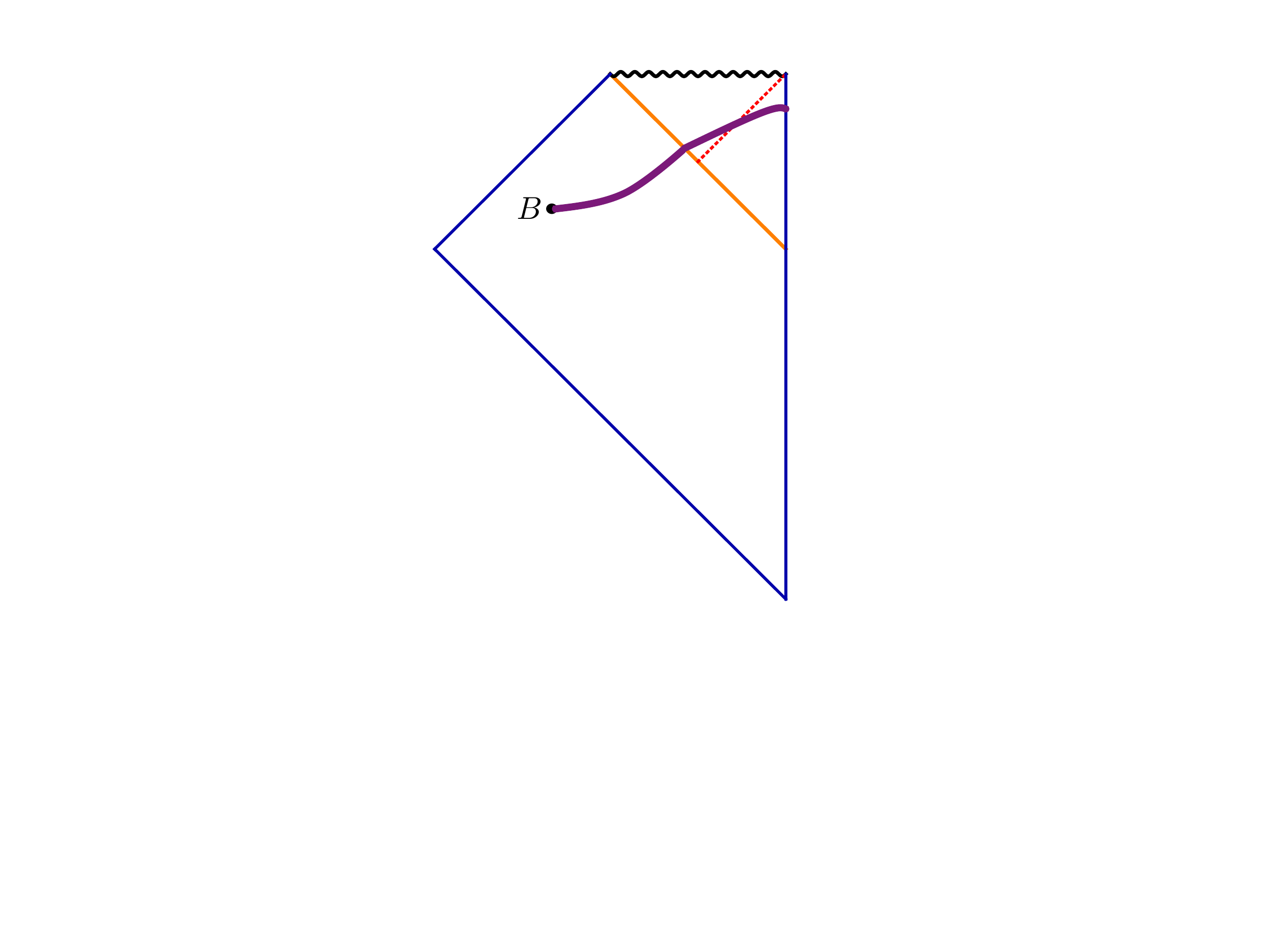}
\end{center}
\caption{
{\it Left}: In the pre-local-equilibration regime, the intersection of the extremal surface with the in-falling shell is close 
to the boundary.
{\it Right}: When $\ft \gtrsim z_h$, the extremal surface starts intersecting with the in-falling shell behind the horizon.
}
\label{fig:surexm}
\end{figure}

For a given $\Sig$, as $\ft$ is varied, the tip of $\Ga_{\Sig}$ traces out a curve $(z_t (\ft), v_t (\ft))$ in a Penrose diagram, which provides a nice way to visualize the evolution of $\Ga_{\Sig}(\ft)$. See Fig.~\ref{fig:curves}. For $\ft \ll \eql$, e.g. point $A$ whose corresponding $\Ga_{\Sig}$ is shown on the left of Fig.~\ref{fig:surexm}, $\Ga_{\Sig}$  
intersects with the in-falling shell close to the boundary.  This defines the pre-local-equilibrium stage discussed in the main text.  When $\ft$ becomes of order $\eql \sim z_h$,  at some point $\Ga_{\Sig}$ starts intersecting the shell behind the horizon, as at point $B$ in Fig.~\ref{fig:curves}, whose corresponding $\Ga_{\Sig}$ is shown on the right of Fig.~\ref{fig:surexm}.

The saturation can proceed as a continuous or discontinuous transition, as illustrated in Fig.~\ref{fig:curves}. For a continuous transition, depicted on the left, the entire curve $(z_t, v_t)$ is single-valued as a function of $\ft$ and saturation happens at point $C$, after which the extremal surface lies entirely in the black hole region. 
In contrast, for a discontinuous transition, depicted on the right of Fig.~\ref{fig:curves}, at saturation the extremal surface jumps from that with tip at point $C'$ to that with tip at $C$.

\begin{figure}[!h]
\begin{center}
\includegraphics[scale=0.6]{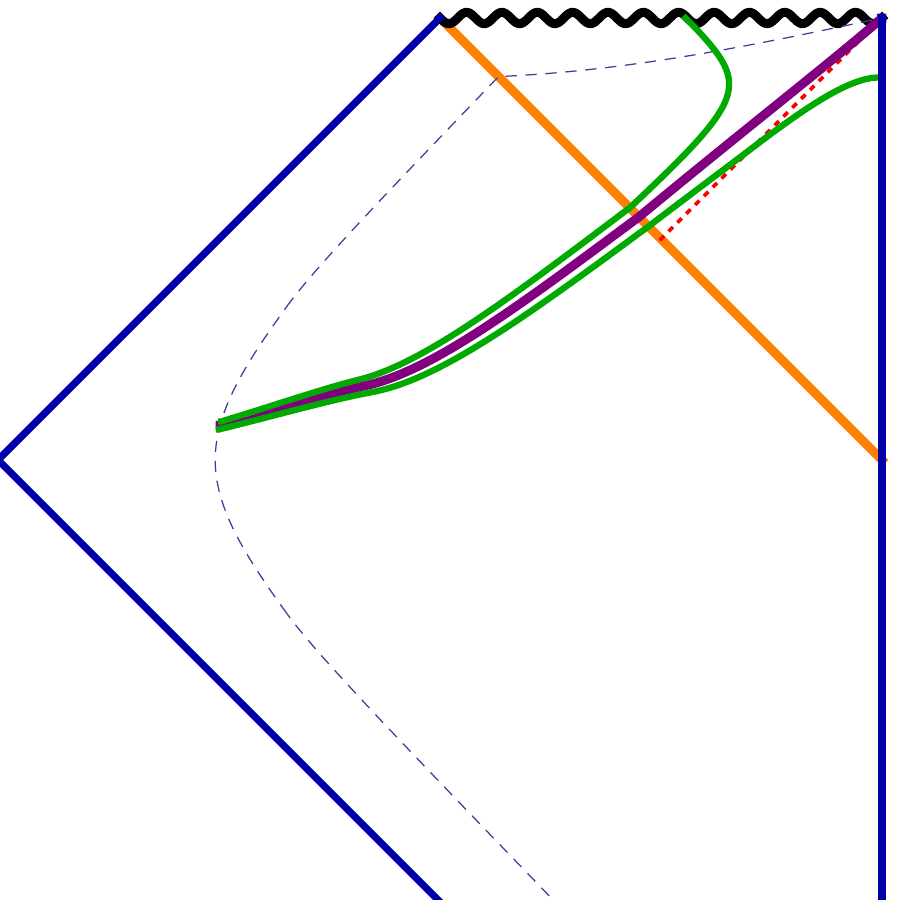} \;
\includegraphics[scale=0.6]{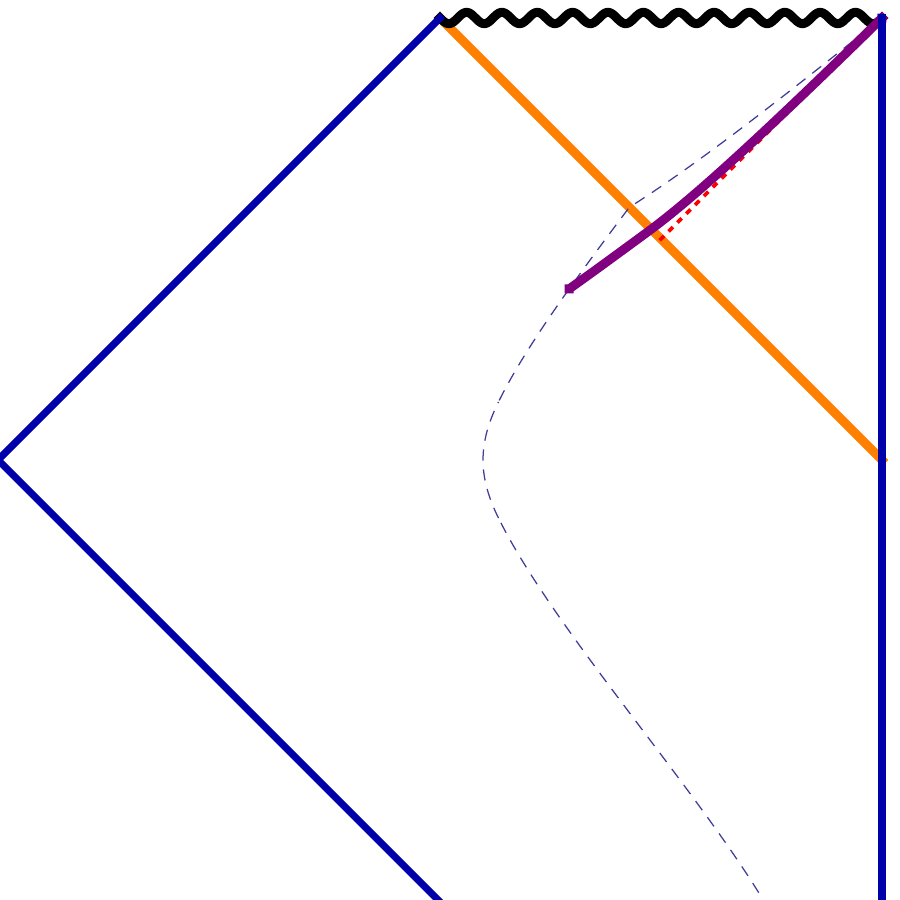}
\end{center}
\caption{
{\it Upper}: The dotted line denotes a curve at constant $z$, along which $v$ increases from $-\infty$ to 
$+\infty$ from bottom (not shown) to top. The purple line corresponds to $\Ga_{\Sig}^*$, the critical extremal surface, while the green lines 
correspond to $\Ga_{\Sig}$ with $v_t$ just above and below $v_t^*$. For $\ft$ in the linear growth regime, 
$z_t (\ft)$ is very large, of order $R$, and the corresponding $\Ga_{\Sig}^*$ asymptotes to a constant $z=z_m$ with $z_m$ determined as described below~\eqref{gene}.
{\it Lower}: At late times $z_t(\ft)$ is $O(1)$ (i.e. does not scale with $R$), and the corresponding $\Ga_{\Sig}^*$ asymptotes to the horizon. This is the regime corresponding to the late time memory loss for a sphere described in~\eqref{evem}. 
}
\label{fig:critical}
\end{figure}

Fig.~\ref{fig:critical} describes the phenomenon of critical extremal surfaces and its relevance for the regimes of linear growth and late-time memory loss. Consider a constant $z_t$ curve
in the Penrose diagram, along which $v_t$ is varied. It turns out there is a critical 
$v_t^* (z_t)$: $\Ga_{\Sig}$ reaches the boundary only for $v_t < v_t^*$, with the {\it critical extremal surface} $\Ga_{\Sig}^*$ corresponding to $v_t^* (z_t)$ stretching to $R, \ft = \infty$. 
For large $R$, $v_t$ is very close to $v_t^* (z_t)$ and $\Ga_\Sig$ closely follows $\Ga_{\Sig}^*$ before deviating from it to reach the boundary.
In the linear growth regime, $\Ga_{\Sig}^*$ asymptotes to $z=z_m > z_h$ which 
is the origin of~\eqref{line} and~\eqref{gene}, while the late-time scaling in \eqref{evem} giving memory loss (which occurs for $\Sig$ a sphere) originates from $\Ga_{\Sig}^*$ asymptoting to the horizon $z=z_h$. 

\section{Appendix B: $v_E$ and $\fR_\Sig$ for various metrics}
 
In this Appendix we discuss the behavior of $\Vee$ and $\fR_\Sig$ for various examples of $h(z)$.

\begin{figure}[!h]
\begin{center}
\includegraphics[scale=0.6]{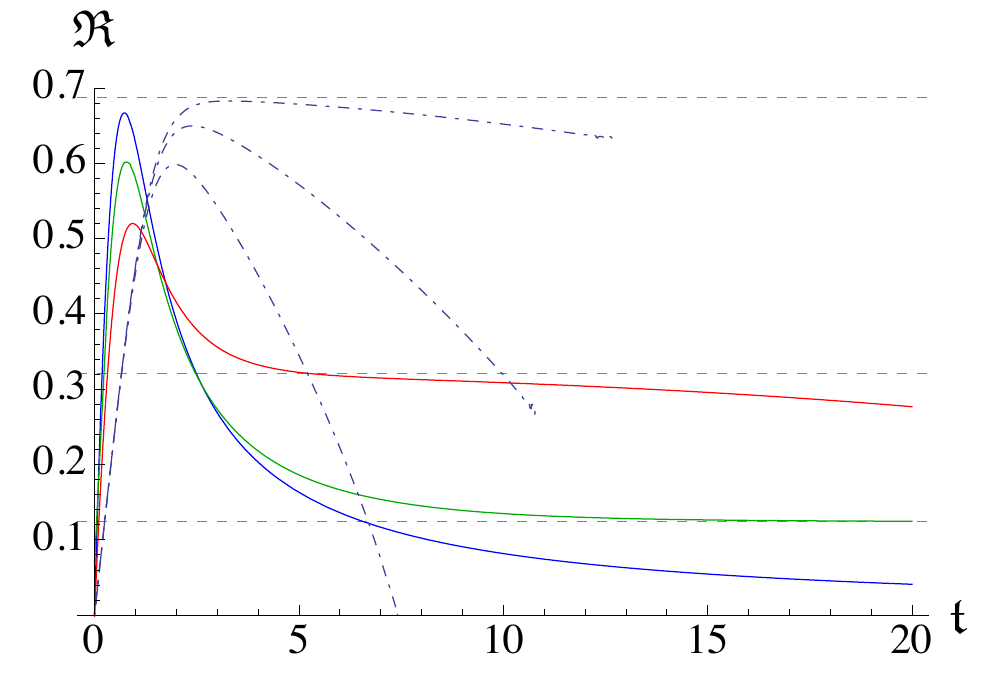} 
\includegraphics[scale=0.6]{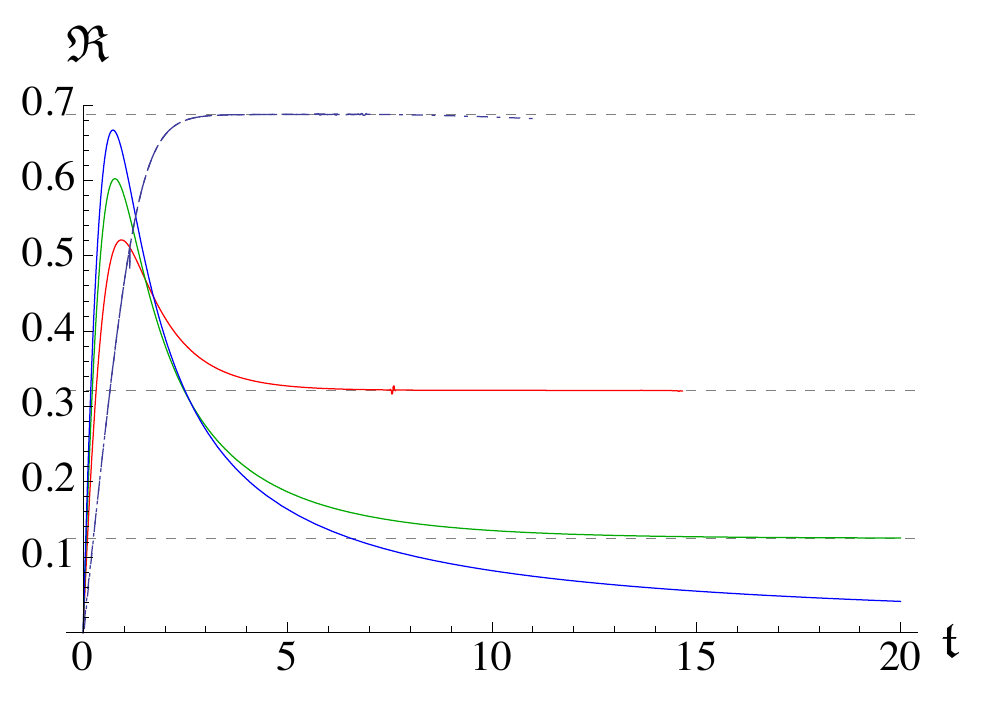}
\includegraphics[scale=0.6]{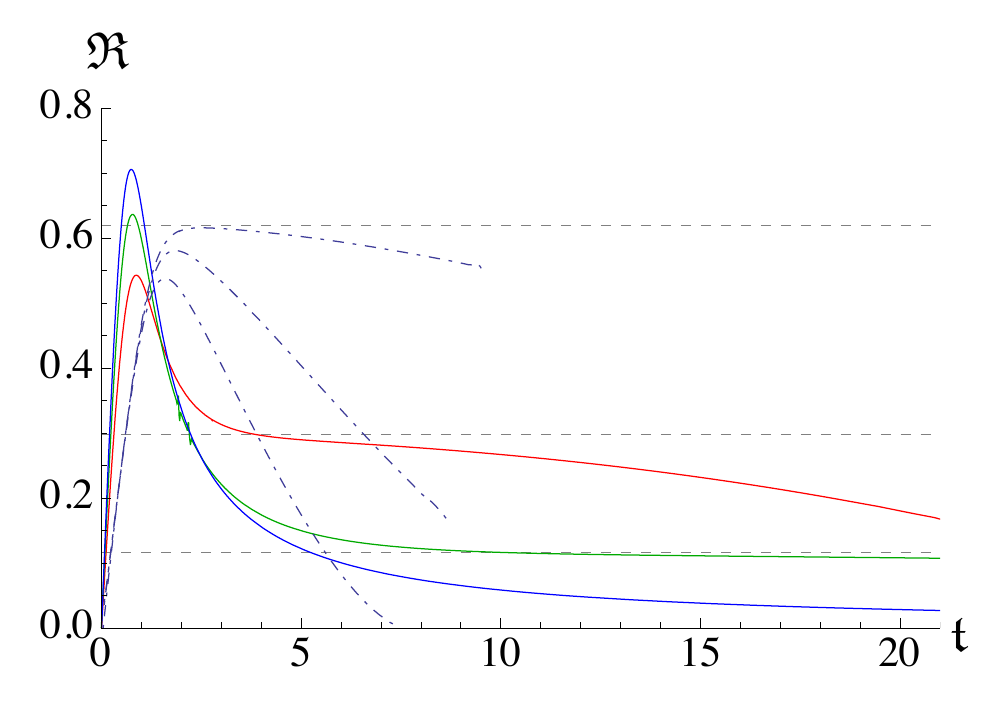}
\end{center}
\caption{$\fR_\Sig$ for $\Sig$ a sphere or strip, for Schwarzschild and RN black holes.
We use units in which the horizon is at $z_h =1$. 
{\it Upper}: For $d=3$ and $\Sig$ a sphere. The dot-dashed curves are for the Schwarzschild black hole with $R = 7$, $13$, and $50$, respectively (larger values of $\ft$ for the $R=13, 50$ curves are not shown due to insufficient numerics), with the top horizontal dashed line marking $\Vee^{(\rm S)}$. 
Red, green, and blue curves are for the RN black hole with $(u =0.5, R=20)$, $(u=0.2, R=50)$, and $(u=0, R=50)$ respectively, and the two lower dashed horizontal lines mark $v_E$ for $u=0.5$ and $0.2$.  
{\it Middle}: For $d=3$ and $\Sig$ a strip. The dot-dashed curves are for the Schwarzschild black hole with $R=7, 12, 15$. It is interesting to note their evolutions are essentially identical with the exception of different saturation times. The visible end of the dot-dashed curves coincides with discontinuous saturation for $R=7$. For $R=12$ and $15$ the curves have not been extended to saturation due to insufficient numerics. 
 The red, green, and blue curves are for the RN black hole with $(u=0.5, R=5)$, $(u=0.5, R=6)$, and $(u=0, R=6)$, respectively. The $u=0.5$ curve ends at saturation, but for $u=0.2$ and $0$, saturation happens at larger values of $\ft$ than shown.
{\it Lower}: For $d=4$ and $\Sig$ a sphere. The color and pattern scheme is identical to the upper plot, but the Schwarzschild curves are at $R=7$, $12$, and $50$, respectively, and $u=0.5$, $0.2$, $0$ curves are all at $R=20$.}\label{fig:frbe}
\end{figure}

Working out the full time dependence of $\fR_\Sig (\ft)$ for large $R$
requires extensive numerical study. We have done so for $\Sig$ a sphere or strip with $h(z)$ given by Schwarzschild and RN black holes. In these cases, we find $\fR_\Sig$ is always bounded by $\Vee^{\rm (S)}$. 
Some plots are presented in Fig.~\ref{fig:frbe}. Note that for a RN black hole, the initial slope 
of $\fR_\Sig (\ft)$ is given by 
\be 
{2 \pi \ov d-1} {\sE \ov s_{\rm eq}} = {1 \ov z_h} \le({d-1 \ov d-2} - \ha {d u \ov d-2} \ri), 
\ee
which is bigger than that for the Schwarzschild black hole value of ${1 \ov 2 z_h}$, while $\Vee$ is smaller, resulting in a peak at time of order 
$\ft \sim \eql\sim z_h$. (Recall $u$ was defined below~\eqref{stvrn} and $u=1$ corresponds to Schwarzschild.) We note that for $d=4$ and sufficiently small $u$, the peak value actually exceeds $\Vee^{(\rm S)}$. 

\begin{figure}[!h]
\begin{center}
\includegraphics[scale=0.5]{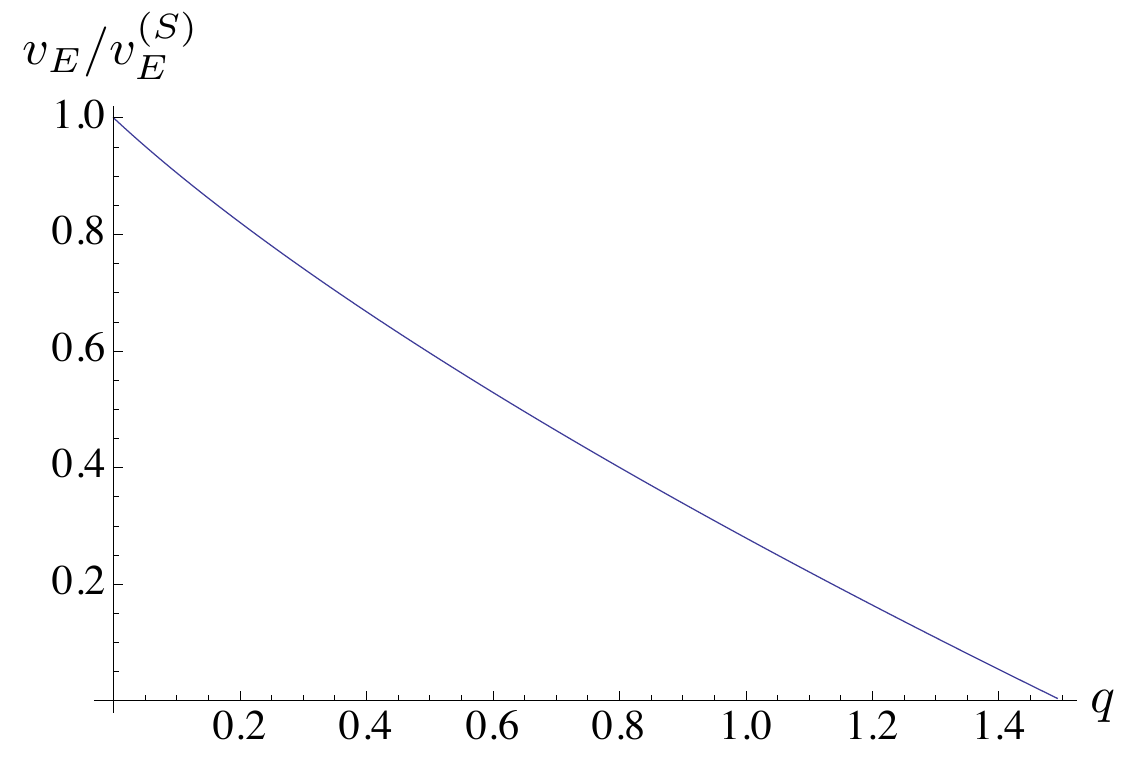} 
\includegraphics[scale=0.7]{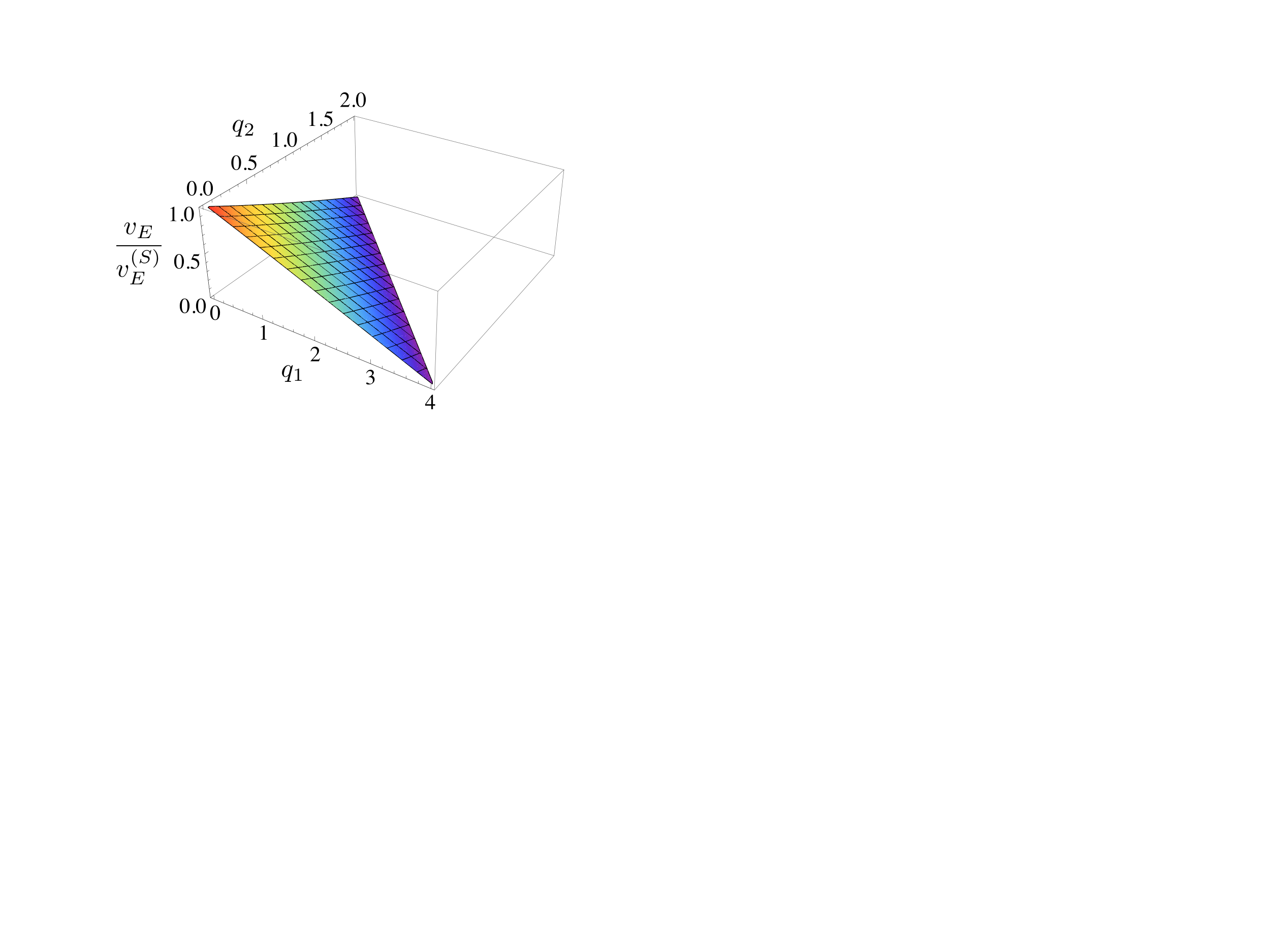} 
\end{center}
\caption{Plots of $v_E/v_E^{(S)}$ in examples of $h(z)$ with parameter space restricted by the NEC and the existence of a horizon. {\it Upper}: For \eqref{REg} with $d=3$ and $p=2$. {\it Lower}: For \eqref{tpe} with $d=4$.
 }\label{fig:ranex1}
\end{figure}

We now consider the behavior of $\Vee$ for more general black holes. Other than Schwarzschild and RN black holes there are no known examples of explicit supergravity solutions of the form~\eqref{BHre1}. 
Given that~\eqref{gene} depends on some location $z=z_m$ behind the horizon, which could be shifted around by modifying $h(z)$, one may naively expect that $\Vee$ could easily be increased by changing $h(z)$ arbitrarily. However, in the examples we studied, the null energy condition 
\be \label{nec}
z^2 h'' - (d-1) z h' \geq 0 \
\ee
constrained $v_E \leq v_E^{\rm (S)}$. These include: 
\bi
\item  A generalization of the RN metric, 
\be \label{REg}
h(z) = 1 - M z^d + q z^{d+p} \ , \qquad p>0\ .
\ee 
The null energy condition~\eqref{nec} requires $q \geq 0$ and in order for the metric to have a horizon (and not a naked singularity), $q  \leq {d \ov p}$. (Here and below we set $z_h=1$). This constrains $v_{E} \leq v_E^{(S)}$, an example of which we show in Fig.~\ref{fig:ranex1}. Note that for $q < 0$, $\Vee$ does exceed $\Vee^{(S)}$. 

\item A three-parameter example with
\be \label{tpe}
h(z) = 1 - M z^d + q_1 z^{d+1} + q_2 z^{d+2} \ ,
\ee
where the null energy condition~\eqref{nec} requires both $q_1$ and $q_2$ to be non-negative, and the existence of a horizon requires $q_1 + 2 q_2 \leq d$. Then again $v_E \leq v_E^{(S)}$, an example of which is shown in Fig.~\ref{fig:ranex1}.
\ei
We have also looked at non-polynomial examples and found $v_E \leq v_E^{\rm (S)}$. The phase space we have explored is certainly tiny, nor do we expect that the null energy condition is the only consistency condition. Nevertheless, the examples are suggestive. 

\subsection{More general supergravity solutions} 

To have more realistic examples requires a generalization of the Vaidya metric~\eqref{vaidya} to 
\be \label{vaidya1}
ds^2 = {L^2 \ov z^2} \le(- f (v,z) dv^2 - 2 k(v, z) dvdz + d \vec x^2 \ri) 
\ee
with $f (v,z) = 1-  \th (v) g(z)$ and $k(v,z) = 1 - \th (v) m(z)$. The black hole part of the spacetime now has a metric of the form 
\be \label{BHre2}
ds^2 = {L^2 \ov z^2} \le(- h (z) dv^2 - 2 n (z) dvdz + d \vec x^2 \ri) 
\ee
with $h(z) \equiv 1 - g(z)$ and $n(z) \equiv 1 - m (z)$, and can also be written as 
\be 
ds^2 = {L^2 \ov z^2} \le(- h (z) dt^2 + {dz^2 \ov l(z)}  + d \vec x^2 \ri) , \;
n^2 (z) = {h(z) \ov l (z)} \ . 
\ee
It turns out that for this more general class of metrics equation~\eqref{gene} still applies with $z_m$ again obtained by minimizing $h(z)/z^{2(d-1)}$. We proceed to consider some explicit examples:

{\it 1. Charged black holes in $\sN =2$ gauged supergravity in AdS$_5$~\cite{Behrndt:1998jd}:}
\be \label{c5}
ds^2={L^2 H^{1 \ov 3}(y) \ov y^2}\le(- h(y)dt^2+ d\vec{x}^2+{dy^2 \ov f(y)}\ri)
\ee
where
\be
h(y) = {f(y) \ov H(y)}, \; f(y)=H(y)  -{\mu y^4}, \;  H(y)=\prod_{i=1}^{3}\le(1+{q_i y^2 }\ri). 
\ee
We normalize $y$ so that the horizon is at $y_h =1$. Then $\mu = \prod_{i=1}^{3}\le(1+q_i \ri)$ and
requiring that the temperature is non-negative gives
\be 
\ka_3 \leq \ka_1 + 2 \ .
\ee
We find 
\be \label{D5ve}
v_E^2 = {2 +  \ka_1 y_m^2 - \ka_3 y_m^6 \ov 1 + \ka_1 + \ka_2 + \ka_3} y_m^{-6}
\ee
with 
\be \label{deek}
\ka_1 = q_1 + q_2 + q_3, \; \ka_2 = q_1 q_2 + q_1 q_3 + q_2 q_3 , 
\; \ka_3 = q_1 q_2 q_3,
\ee
and 
\be 
y_m^2 = {\ka_1 + \sqrt{\ka_1^2 + 3 (1 + \ka_1 + \ka_3)} \ov 1 + \ka_1 + \ka_3} \ .
\ee
It can be readily checked analytically that for one- and two-charge cases with $q_3=\ka_3 =0$, the bound is satisfied for any $(q_1,q_2)$, including regions which are thermodynamically unstable. After numerical scanning we find that \eqref{D5ve} satisfies $v_E \leq v_E^{(S)}$ in the full three-charge parameter space.

{\it 2. Charged black holes in $\sN =8$ gauged supergravity in AdS$_4$~\cite{Duff:1999gh}:}
\be\label{c4}
ds^2={L^2 H^\ha (y) \ov y^2}\le(-h(y)dt^2+ d\vec{x}^2+{dy^2 \ov f(y)}\ri)\ ,
\ee
where
\be
h(y) = {f(y) \ov H(y)} , \; f(y)=H(y)  -{\mu y^3 }, \; H(y)=\prod_{i=1}^{4}\le(1+{q_i y }\ri) .
\ee
We again set $y_h=1$. Then $\mu = \prod_{i=1}^{4}\le(1+q_i \ri)$ and
requiring non-negative temperature gives
\be
\ka_4 \leq 2 \ka_1+\ka_2+3 \ .
\ee
 We then find thay
\be 
v_E^2 = {3 + 2 \ka_1 y_m +\ka_2 y^2_m - \ka_4 y_m^4 \ov 1 + \ka_1 + \ka_2 + \ka_3 + \ka_4} y_m^{-4}
\ee
where $\ka_i$ are defined analogously to~\eqref{deek}, with e.g. $\ka_4 = q_1 q_2 q_3 q_4$, and $y_m$ is the smallest positive root of the equation
\be
(1+\ka_1+\ka_2 + \ka_4) y^3 - 2 \ka_2 y^2 - 3 \ka_1 y - 4 =0 \ .
\ee
It can again be readily checked that for a single charge $q_1 \neq 0$ $v_E \leq v_E^{(S)}$ is satisfied for any $q_1$. One finds after numerical scanning that the bound is in fact satisfied in the full four-parameter space.

{\it 3. Metrics with hyperscaling violation:} Finally, let us consider metrics with hyperscaling violation and dynamical exponent $1$~\cite{Gouteraux:2011ce,Dong:2012se},
\be\label{hysv}
ds^2={L^2 \ov y^2}\le({y \ov y_F}\ri)^{2 \th \ov d-1} \le(- f(y)dt^2+{dy^2 \ov f(y)}+d\vec{x}^2\ri)
\ee
where $f(y)=1-\le({y \ov y_h}\ri)^{\tilde d }$ and $\tilde d \equiv d - \th$. $y_F$ is some scale and $\th$ is a constant. Dimensionally reduced near-horizon Dp-brane spacetimes (for which $d=p+1$) are examples with 
$\th = - {(d-4)^2 \ov 6-d}$. With boundary at $y=0$, such metrics are no longer asymptotically AdS, but our discussion can still be applied. 
For~\eqref{hysv} we find 
\be 
v_E^2 = {\le(\tilde \eta-1\ri)^{\tilde \eta-1} \ov \tilde \eta^{\tilde \eta}}\ , \qquad 
\tilde \eta ={2 (\tilde d -1) \ov \tilde{d}}\ .
\ee
The null energy condition now reads~\cite{Dong:2012se}  
\be
\tilde d \th \leq 0
\ee 
which implies either $\th \leq 0$ or $\tilde d \leq 0$. The former leads to $\tilde d \geq d$ and thus $v_E \leq v_E^{(\rm S)}$, while the latter is inconsistent with small $y$ describing UV physics. For examples coming from Dp-branes, $\th$ is clearly negative with $d \leq 6$, while for higher $d$ the metric no longer describes a non-gravitational field theory.

\end{document}